\documentclass[sigconf]{acmart}

\usepackage{enumitem}
\usepackage{xcolor}
\usepackage{subcaption} 
\usepackage{microtype}                 
\usepackage{textcomp} 
\usepackage{acmart-taps}

\usepackage{listings}
\usepackage{color}

\definecolor{dkgreen}{rgb}{0,0.6,0}
\definecolor{hannah}{rgb}{0, 0.5, 0.5}
\definecolor{gray}{rgb}{0.5,0.5,0.5}
\definecolor{mauve}{rgb}{0.58,0,0.82}

\definecolor{lightgray}{rgb}{.9,.9,.9}
\definecolor{darkgray}{rgb}{.4,.4,.4}
\definecolor{purple}{rgb}{0.65, 0.12, 0.82}

\lstdefinelanguage{JavaScript}{
  keywords={typeof, new, true, false, catch, function, return, null, catch, switch, var, if, in, while, do, else, case, break, d},
  keywordstyle=\color{blue}\bfseries,
  ndkeywords={class, export, boolean, throw, implements, import, this},
  ndkeywordstyle=\color{darkgray}\bfseries,
  identifierstyle=\color{black},
  sensitive=false,
  comment=[l]{//},
  morecomment=[s]{/*}{*/},
  commentstyle=\color{purple}\ttfamily,
  stringstyle=\color{red}\ttfamily,
  morestring=[b]',
  morestring=[b]"
}

\AtBeginDocument{%
  }

\copyrightyear{2023} 
\acmYear{2023} 
\setcopyright{rightsretained} 
\acmConference[IUI '23]{28th International Conference on Intelligent User Interfaces}{March 27--31, 2023}{Sydney, NSW, Australia}
\acmBooktitle{28th International Conference on Intelligent User Interfaces (IUI '23), March 27--31, 2023, Sydney, NSW, Australia}
\acmDOI{10.1145/3581641.3584041}
\acmISBN{979-8-4007-0106-1/23/03}

\begin{document}

\title{User-Driven Support for Visualization Prototyping in D3}

\author{Hannah K. Bako}
\email{hbako@cs.umd.edu}
\orcid{0000-0001-7958-1319}
\affiliation{%
  \institution{University of Maryland}
  \streetaddress{}
  \city{College Park}
  \state{MD}
  \country{USA}
  \postcode{20774}
}

\author{Alisha Varma}
\email{alishav@umd.edu}
\affiliation{%
  \institution{University of Maryland}
  \streetaddress{}
  \city{College Park}
  \state{MD}
  \country{USA}
  \postcode{20774}
}

\author{Anuoluwapo Faboro}
\email{afaboro@umd.edu}
\affiliation{%
  \institution{University of Maryland}
  \streetaddress{}
  \city{College Park}
  \state{MD}
  \country{USA}
  \postcode{20774}
}

\author{Mahreen Haider}
\email{mhaider1@umd.edu}
\affiliation{%
  \institution{University of Maryland}
  \streetaddress{}
  \city{College Park}
  \state{MD}
  \country{USA}
  \postcode{20774}
}

\author{Favour Nerrise}
\email{fnerrise@umd.edu}
\affiliation{%
  \institution{University of Maryland}
  \streetaddress{}
  \city{College Park}
  \state{MD}
  \country{USA}
  \postcode{20774}
}

\author{Bissaka Kenah}
\email{bkenah@umd.edu}
\affiliation{%
  \institution{University of Maryland}
  \streetaddress{}
  \city{College Park}
  \state{MD}
  \country{USA}
  \postcode{20774}
}

\author{John P. Dickerson}
\email{john@cs.umd.edu}
\affiliation{%
  \institution{University of Maryland}
  \streetaddress{}
  \city{College Park}
  \state{MD}
  \country{USA}
  \postcode{20774}
}

\author{Leilani Battle}
\email{leibatt@cs.washington.edu}
\affiliation{%
  \institution{University of Washington}
  \streetaddress{}
  \city{Seattle}
  \state{WA}
  \country{USA}
  \postcode{}
}

\renewcommand{\shortauthors}{Bako et al.}

\begin{abstract}
  Templates have emerged as an effective approach to simplifying the visualization design and programming process. For example, they enable users to quickly generate multiple visualization designs even when using complex toolkits like D3. However, these templates are often treated as rigid artifacts that respond poorly to changes made outside of the template's established parameters, limiting user creativity. Preserving the user's creative flow requires a more dynamic approach to template-based visualization design, where tools can respond gracefully to users' edits when they modify templates in unexpected ways. 
In this paper, we leverage the \emph{structural similarities} revealed by templates to design resilient support features for prototyping D3 visualizations: \textbf{recommendations} to suggest complementary interactions for a users' D3 program; and \textbf{code augmentation} to implement recommended interactions with a single click, even when users deviate from pre-defined templates. We demonstrate the utility of these features in Mirny, a design-focused prototyping environment for D3. In a user study with 20 D3 users, we find that these automated features enable participants to prototype their design ideas with significantly fewer programming iterations. We also characterize key modification strategies used by participants to customize D3 templates. Informed by our findings and participants' feedback, we discuss the key implications of the use of templates for interleaving visualization programming and design. 

\end{abstract}




\maketitle

\section{Introduction}
Creating thoughtful and captivating data visualizations necessitates careful consideration of  multiple design alternatives as designers program their visualizations. This process often requires an iterative exercise of refining one or more design options until an optimal visualization design is achieved i.e., \textit{visualization prototyping}~\cite{walny2019data, bako2022understanding, bigelow2014reflections, mckenna2014design}.

This paper explores the interplay between \emph{visualization programming}, i.e., creating interactive visualizations via languages, and \emph{visualization design}, i.e., creating visualizations via design tools, in supporting the \emph{visualization prototyping} process.
With visualization programming, a user can create dozens of interactive visualizations just by permuting a few core variables within a single program. We see these benefits in well-designed languages like D3 that have thousands of visualization programs observed online~\cite{battle2022exploring,battle2018beagle}. However, users must first be knowledgeable of the underlying language, which can be a huge hurdle for individuals with limited or no programming experience~\cite{battle2022exploring,satyanarayan2019critical, walny2019data}. For example, even though interactions are considered essential to defining information visualization~\cite{card1999using}, interactive components such as brush filters, panning and zooming, etc., are notoriously difficult to implement in D3 even for experienced users~\cite{satyanarayan2017vega-lite,zong2020lyra}.
In contrast to visualization programming, visualization design encourages users to explore how to visualize data using visual elements which may be outside the bounds of traditional programming contexts~\cite{wun2019you}. For example, Data Illustrator~\cite{liu2018dataIll} and Lyra/Lyra2~\cite{satyanarayan2014lyra,zong2020lyra} enable users to quickly create highly customized and detailed visualizations. That being said, the process of evaluating multiple alternative designs can still be an arduous task in design tools since the user must edit the designs themselves to generate alternatives.

Visualization programming and visualization design contribute complementary yet distinct benefits to the visualization prototyping process that users struggle to achieve with only one or the other. Particularly, it is difficult to balance out the programming effort needed to explore the space of possible visualization designs \emph{and} accompanying interactions. The difficulty of maintaining this balance points to a core question that drives our work: How can we make it easier for users to prototype their interactive design ideas through complex visualization languages like D3?

One solution is to abstract visualization examples into general-purpose programming templates~\cite{McNutt2021Templates,harper2018convertingd3}, and generate code recommendations based on these pre-defined structures (e.g. Falx~\cite{wang2021falx}).
In prior work, programming templates are often treated as pre-defined \emph{static} structures, where users can only modify fixed design parameters to create visualizations. However, users may be interested in changing multiple parts of a template such as aesthetics (e.g., color schemes), data encodings, interaction styles (e.g., slider versus brush filters), and mark types. Moreover, these parameters are chosen by the tool developers, not their users. Additionally, prior work finds that designers often dissect existing examples to extract new functionality they want to build on~\cite{battle2022exploring, bako2022understanding}; for example, users could start working with a bar chart example but eventually transform it into a completely different visualization. By limiting how users can modify a template, we hamper these creative design activities.

\begin{figure*}[tbp]
    \centering
   \includegraphics[width=0.9\linewidth,  trim=1cm 2cm 1cm 2cm]{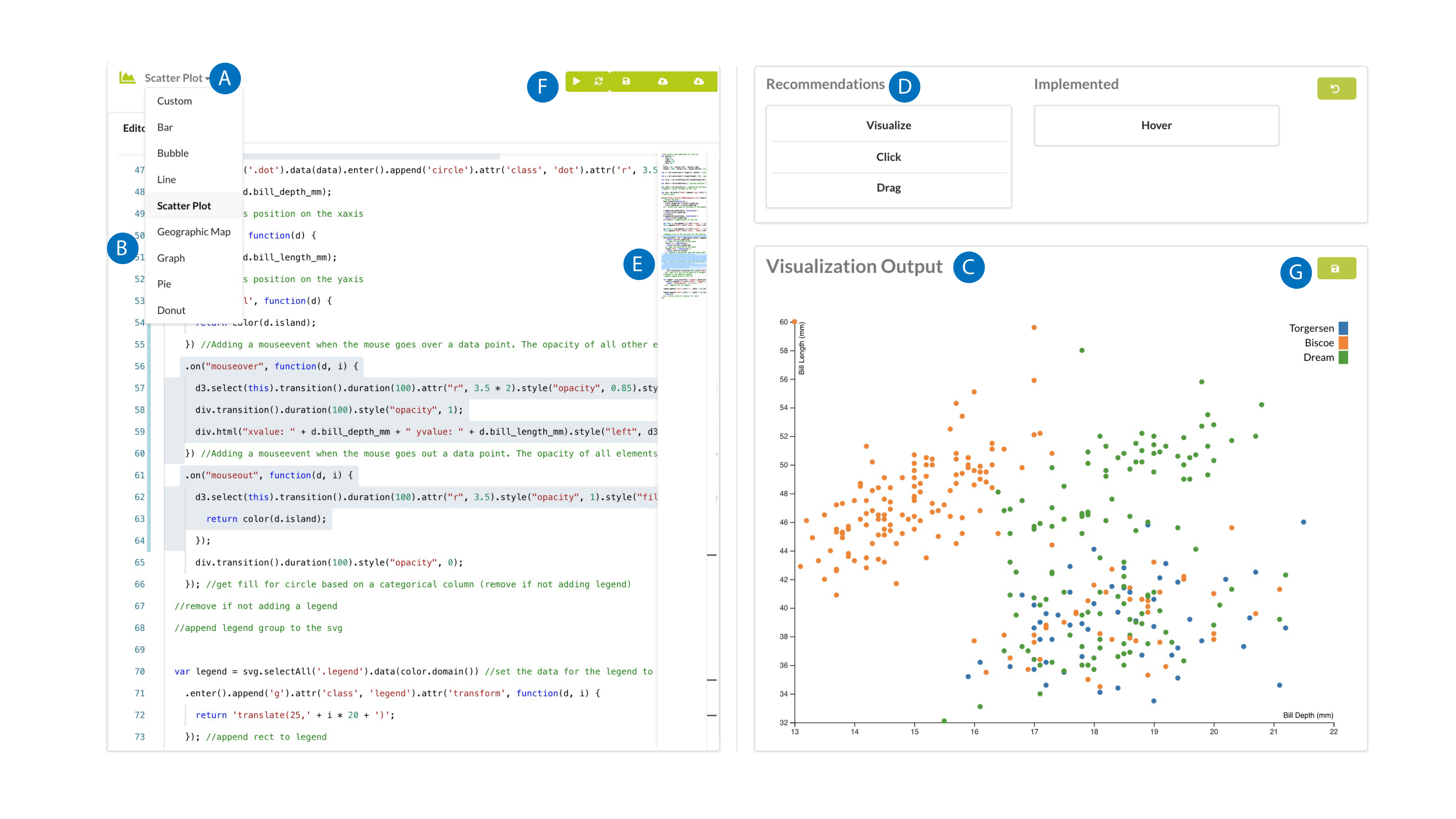}
   \vspace{-5mm}
  \caption{An example of an interactive visualization designed using Mirny. A user can select a visualization such as a \textit{scatterplot} to create from the \textbf{Template Panel (A)}. The code responsible for creating the \textit{scatterplot} is displayed within the \textbf{Editor (B)} and the corresponding visualization is rendered in the \textbf{Visualization Panel (C)}. A list of recommended interactions is displayed to the user in the\textbf{ Recommendation Panel (D)}. A user can then choose a recommended interaction such as a \textit{Hover} and the code responsible for implementing the interaction is added to the interface and highlighted for the user to view and modify \textbf{(E)}. Once a user is satisfied with their visualization, they can export the code and visualization using the \textbf{Controls (F)} and \textbf{(G)}.}
  \Description{A snapshot of the Mirny user interface containing a code editor with highlighted D3 code on the left half of the page, the recommendation panel at the top right corner, and the visualization panel at the bottom right corner. The template panel drop-down is also visible with the scatterplot template highlighted. Each component is labeled.}
  \label{fig:teaser}
\end{figure*}

In this paper, we propose a \emph{dynamic} approach to template-based design for interactive visualizations.
Rather than restricting how templates can be modified, our goal is to identify the \emph{design patterns} encapsulated within these templates, and develop automated features to identify and extend these patterns as D3 users program in real-time. 
These features must be \emph{resilient}, meaning they should work regardless of whether the user is populating template code or writing code from scratch. To achieve this, we leverage a suite of D3 templates derived in our prior work~\cite{bako2022streamlining} to build two automated support features for D3: (1) a user-driven recommendation model that suggests suitable interaction[s] to add to a user's visualization code and (2) a code augmentation feature that can integrate code snippets directly into live user code. 
In this way, we can help users \emph{preserve their creative design thinking} as they code, and work towards closing the gap between visualization design and visualization programming.
To demonstrate the utility of these features, we implemented them in Mirny, a design-focused environment for rapid prototyping of interactive visualizations in D3 (shown in~\autoref{fig:teaser}).

We evaluated these features
through a study with 20 D3 users prototyping interactive visualizations in Mirny. We find that using the automated features implemented in Mirny, participants were able to prototype design ideas in less time and with fewer programming iterations compared to their typical D3 implementation workflows.
We observed participants modifying templates in a variety of interesting ways, which we synthesize into three high-level modification strategies:
changing the visual styles (e.g. mark colors or mark types), introducing data transformations (e.g., to aggregate or sort data), and removing or adding new data encodings. 

In summary, this paper makes the following contributions:

\begin{itemize}[nosep]
\item We present a user-driven \textit{model for recommending interactions} users can  implement within a D3 prototype 
\item We develop a code augmentation feature to \textit{automatically integrate interaction templates} into live D3 code. 
\item We implement these features in Mirny and evaluate Mirny through \textit{a user study with 20 D3 users}. Our results show Mirny enabled participants to implement interactive D3 visualizations in less time and with fewer iterations.
\item We synthesize \textit{three high-level modification strategies} used by designers to customize visualization templates.
\end{itemize}

Using our findings, we suggest research opportunities to enhance template-based design approaches within visualization languages. All our data and code are shared on OSF~\footnote{\textcolor{blue}{\url{https://osf.io/97mru/?view_only=c89bc8b4f1a64799ad4453213a255744}}}
\section{Related Work} \label{sec:related-works}

Our work draws from research in 1) visualization authoring and recommendations, 2) interactive visualizations, and 3) template and code generation. We review the relevant literature for each topic below.

\subsection{Visualization Authoring and Recommendation Tools}
There are a plethora of visualization authoring tools available to users such as specification-based languages (e.g~\cite{plotly2015,bostock2011d3,satyanarayan2016vega,satyanarayan2016reactivevega}), automated tools (e.g~\cite{stolte2002polaris, satyanarayan2014lyra, zong2020lyra}), or graphic-driven tools that rely on GUI widgets (e.g~\cite{liu2018dataIll,zong2020lyra}). Specification languages are expressive but require considerable programming skills to use~\cite{vartak2017towards, dibia2019data2vis, hu2019vizml}. On the other hand, graphic-driven tools eliminate the need for programming. However, by eliminating the need for specifications, they take away the control from the user~\cite{horvitz1999principles} and still have a learning curve~\cite{satyanarayan2019critical}. 

To help users navigate the space of design options, recommendation tools generate design alternatives for users to choose from, rather than have users create alternatives from scratch~\cite{wongsuphasawat2016voyager, wongsuphasawat2017voyager,mackinlay2007showme, zeng2021evaluation, luo2018deepeye,moritz2018draco, hu2019vizml,yan2020autosuggest, dibia2019data2vis}. More recently, research has explored the space of generative design tools which allow users to generate and evaluate design alternatives for glyphs~\cite{brehmer2021generative} and dashboards~\cite{ma2020ladv}. Some have explored the data-agnostic  design of visualization prototypes, eliminating the data constraints that designers often face when coming up with visualization designs~\cite{tsandilas2020structgraphics, vuillemot2017structuring}.

However, these systems do not account for current practices that visualization designers engage in when iterating over visualization designs. Recent work finds that when designers iterate over visualization designs, they often integrate multiple design ideas~\cite{bako2022understanding} or integrate code fragments from multiple examples into a single design~\cite{bako2022streamlining, battle2022exploring}. In this regard, we deviate from prior work by shifting our focus from the design of static visualizations from a single template to instead exploring how we can support the integration of multiple code templates during visualization prototyping.

\subsection{Interactive Visualizations}
Interactions are critical for helping users gain a deeper understanding of a dataset through manipulation~\cite{dimara2020interactions, Saket2017Paradigm,yi2007roleofinteractions}. However, visualization toolkits like Chart.js~\footnote{A simple Javascript library for creating charts (\textcolor{blue}{https://www.chartjs.org/})} and recommendation systems generally focus on variation in visual encodings and only provide default interactions such as tooltips and clicks~\cite{mackinlay2007showme, wongsuphasawat2016voyager}. Interaction+~\cite{lu2017interaction+} addresses this challenge by extracting information about visual objects from web pages and using this information to provide a variety of interactions for manipulating the existing visualization. However, the interaction implementation process in Interaction+ is still abstracted from users.
Lyra 2 allows users to specify interactions through demonstration~\cite{zong2020lyra} but hides the underlying specification. Additionally, the space of appropriate interactions for a visualization design is vast~\cite{bako2022streamlining, yi2007roleofinteractions} and users often find it hard to identify what interactions to include in their designs and how to implement these interactions~\cite{battle2022exploring}. Therefore, we observe a need to support rapid exploration and specification of interactions for complex languages like D3.

We aim to help users design and prototype interactive visualizations by giving users access to the actual code. 
To do this, we need to model the space of possible interaction designs, which is often represented through interaction task taxonomies; we defer to existing surveys for more details, e.g., \cite{gathani2022grammar,brehmer2013typology}. In our previous work, we developed a taxonomy of D3 interactions~\cite{bako2022streamlining} based on a subset of tasks from Brehmer and Munzner's  typology~\cite{brehmer2013typology} to model the space of interactions a user may include in their visualizations.

\subsection{Templates and Code Generation}
General code reuse through templates has been advocated for~\cite{kim2004ethnographic, lin2017mining} and leveraged to enable semantic-based code search and code completion~\cite{Reiss2009semantics}.  Work such as Reiss's code search approach~\cite{Reiss2009semantics} has informed the development of tools for finding relevant code snippets (e.g.,~\cite{Gu2018deepcodesearch, Ke2015reparing, McMillan2011portfolio}) and recommending significant code templates to complete a user's program (e.g.,~\cite{Raychev2014codecompletion, Nguyen2012graphcodecompletion, Ponzanelli2014stackoverflow}). Within the visualization community, certain tools use templates to generate visualization designs~\cite{McNutt2021Templates, Mauri2017RawGraphs} and styles~\cite{harper2018convertingd3}. For example, McNutt et al. use parameterized declarative templates as a means of abstracting declarative visualization grammars to promote reuse, exploration, and simplification of the visualization programming process~\cite{McNutt2021Templates}. Templates have also been used in tools like Wrex~\cite{drosos2020wrex}, Falx~\cite{wang2021falx}, and Mage~\cite{Kery2020mage} to provide readable code in response to users' interactions with GUI widgets. However, current work focuses on templates as static artifacts that offer limited control to users. We introduce a \emph{dynamic} approach that detects template-like patterns within a user's visualization program and can integrate design recommendations automatically, even when the user writes code from scratch.

\section{Motivating Usage Scenario for Mirny}
\label{sec:motivating_scenario}
Sandra is a student who wants to create some interesting visualizations for the final project in her data science course. Although this course was broad in scope, it lacked depth in terms of learning interactive visualization techniques.
Sandra has heard that D3 can support a wide variety of interactive visualizations. Although she has limited experience using D3, Sandra wants to create a D3 visualization, preferably with a simple dataset to start, such as the palmerpenguins dataset~\cite{gorman2014ecological}. Sandra decides to use Mirny, shown in \autoref{fig:teaser}, to program her visualizations. Sandra loads Mirny and imports her dataset, she chooses a Scatterplot from Mirny's Template Panel (~\autoref{fig:teaser}a). Mirny automatically generates the D3 code for a Scatterplot from the data as shown in (~\autoref{fig:teaser}b), and the resulting visualization is rendered in the output panel (~\autoref{fig:teaser}c). Mirny recommends a ranked list of common interactions for Scatterplots (\autoref{fig:teaser}d).

Sandra sees nothing wrong with the scatterplot, but she wonders what the distribution of penguins is across the three islands recorded in the dataset. For her second prototype, she selects a bar chart from Mirny's Template Panel.  Mirny automatically generates the D3 code for a bar chart a populates the Editor (~\autoref{fig:teaser}a), saving Sandra the effort of figuring out how to change her design to a different visualization type. Sandra edits the code to use the island attribute for the x-axis and the count of penguins for the Y-axis. The overall visualization looks okay, but Sandra still isn't satisfied. She navigates back her Scatterplot prototype using the Template Panel and changes the color encoding from the species attribute to the island attribute.

Sandra likes the current visualization design but wants to add an interaction. Sandra clicks on the first item from the recommendation panel  (\textit{hover}) and Mirny automatically extends her current D3 program with the corresponding code, highlighting the new code block and providing a summary of its behavior in the Editor (~\autoref{fig:teaser}e). The new code displays additional information about a data point when a hover event is triggered. However, Sandra notices that some of the data points are really close to each other. She also adds a zoom interaction from the recommendation panel, so her viewers can zoom into tightly packed sections of the visualization. Once Sandra completes her visualization prototypes, she exports the code from the editor as a JS file and saves the visualization output in SVG format.

Using Mirny, a D3 user like Sandra can quickly prototype multiple interactive visualizations. Mirny eases the cumbersome process of programming D3 code, enabling users to explore visualization design alternatives using a single interactive interface. Although this example focuses on a new D3 user (Sandra), it can also easily apply to a more advanced user. For example, a professional analyst who has used D3 in several projects may find it tedious to edit an old example from a previous project, and may still prefer to fill in one of Mirny's templates instead. We discuss the experiences of both new and more advanced D3 users in~\autoref{sec:results}.

\section{Preliminary Study: Lessons from Analyzing D3 Examples Online}
\label{sec:prelim}

\begin{figure*}[tbp]
\begin{subfigure}{0.6\linewidth}
    \centering
    \includegraphics[width=0.9\columnwidth, clip]{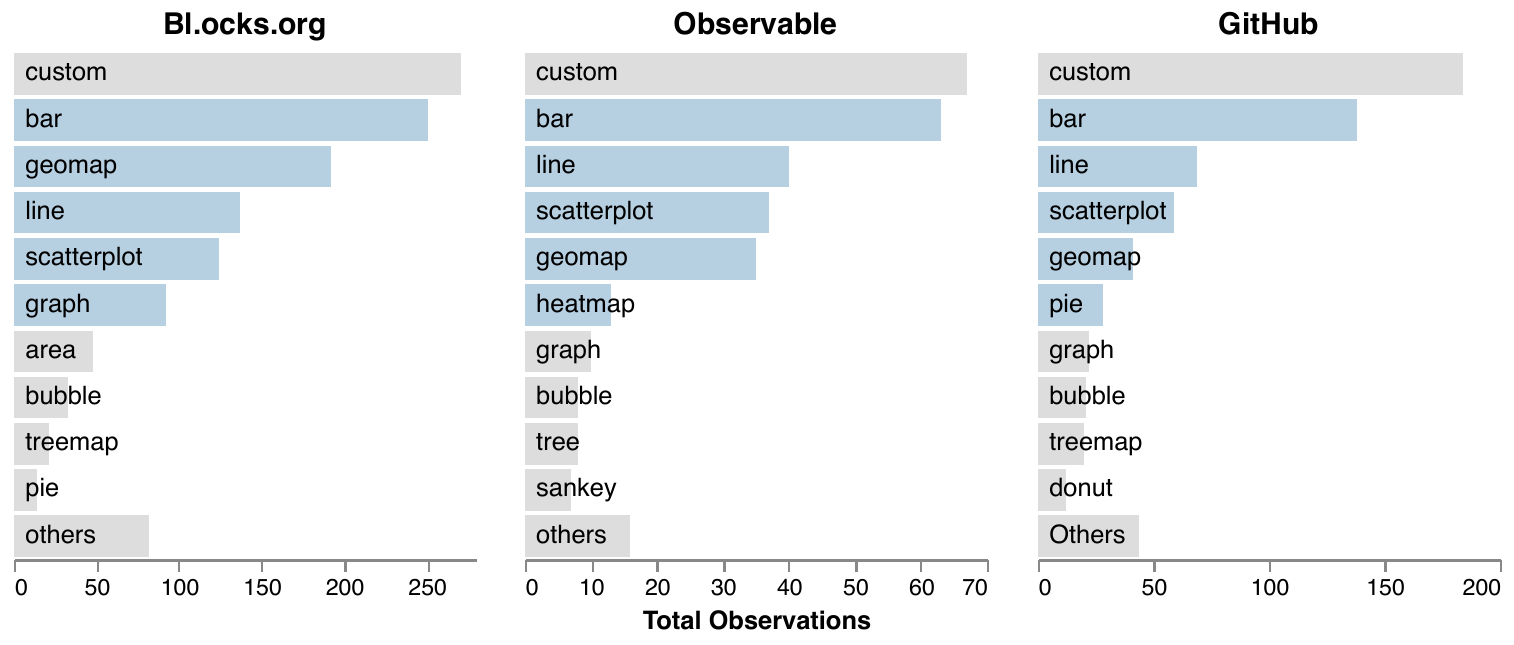}
    \caption{Rankings of observed visualizations in our analysis.}
    \Description{A figure made up of 3 bar charts arranged in a horizontal layout. Each bar chart represents the bl.ocks.org, observable, and GitHub repositories respectively. On the y-axis are the visualization types and on the x-axis are the total recorded observations in our corpus. The top 5 most popular bar charts are colored blue while the remaining bar charts are grey.}
    \label{fig:viz_types}
\end{subfigure}%
\begin{subfigure}{0.4\linewidth}
    \centering
    \includegraphics[width=0.9\columnwidth, clip]{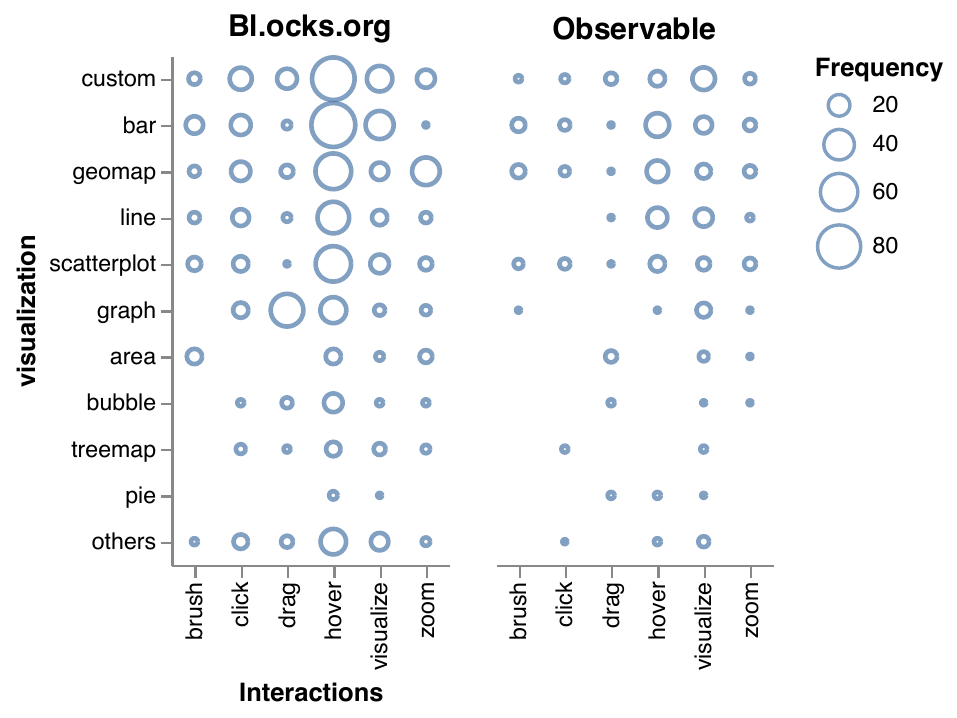}
    \vspace{-3mm}
    \caption{Observed visualization-interaction pairings.}
    \Description{A figure with two scatterplots arranged in a horizontal layout. Each bar chart represents the bl.ocks.org, and observable repositories respectively. On the y-axis are the visualization types and on the x-axis are the interactions. The size of the circles encodes the number of times each interaction was implemented in a specific visualization.}
    \label{fig:int_types}
\end{subfigure}%
\vspace{-2mm}
\caption{Results from the analysis carried out on the three example corpus~\cite{bako2022streamlining} (a) presents the frequency of the top visualization types and their corresponding interactions observed in our datasets. Visualizations with 1\%  and less frequency are captured in the \textit{"others"} category. 
(b) shows the number of times different interaction types were implemented for each visualization type in our corpus.}
\label{fig:data_results}
\vspace{-3mm}
\end{figure*}

The goal of our work is to streamline the process of prototyping visualization designs with complex toolkits like D3. To do this, we first need to understand \emph{what} kinds of visualizations D3 users create and \emph{how} these users implement these visualizations in D3. In our prior work~\cite{bako2022streamlining}, we qualitatively analyzed 2500 D3 examples shared on GitHub~\cite{GitHub:2008}, Bl.ocks.org~\cite{blocks:2016} and Observable~\cite{observable:2016}. Through this analysis, we investigated common implementation patterns D3 users adopt when creating interactive visualizations. In this section, we summarize the key results from this analysis and describe how these results influence design considerations for the Mirny system.

\noindent \textbf{Takeaway 1: A Small Subset of Visualizations are Commonly Implemented in D3.}
To better understand which visualizations tend to be implemented in D3, we counted the incidence of visualization types across our corpus using Battle et al.'s D3 visualization taxonomy~\cite{bako2022streamlining, battle2018beagle}. Our observations reveal that 80\% of the D3 visualizations shared online can be accounted for by just five visualization types (see~\autoref{fig:viz_types}): Bar charts, Geomaps, Line charts, Scatterplots, and Force Directed Graphs. Our findings suggest that \textit{a small subset of visualizations cover the majority of D3 users' needs}.

\noindent \textbf{Takeaway 2: D3 Implementation Patterns are Consistent Within Visualization Types.}
In a subsequent round of analysis, we divided our corpus by visualization type and analyzed the code for similarities and differences across examples, such as differences in the rendering strategy or differences in which interactions were implemented~\cite{bako2022streamlining}. We found that in most cases, \textit{the underlying structure of the D3 code examples did not vary}. In other words, the underlying structure and API calls remained the same within each visualization type. As a result, we observe clear patterns in how different D3 users implement the same types of visualizations, with  variations made to suit users' design preferences. 
We also found that D3 users copy code from existing examples, and even attribute the original sources in their visualization programs, corroborating observances of code reuse in the D3 community by prior work~\cite{battle2022exploring}. To make our findings actionable for the visualization community, we curated and shared templates for 8 popular D3 visualization typest~\footnote{ The curated templates can be found at \label{osf-short} \textcolor{blue}{\url{https://osf.io/k58bp/?view_only=72fa3798bbaa4263b5ad662b26a70cb3}}}. 

\noindent\textbf{Takeaway 3: Interactions are Commonly Implemented in D3 and Vary by Visualization Type.} Interactions are known to be challenging to implement in D3~\cite{satyanarayan2017vega-lite,battle2022exploring}, and thus are a promising focus for design templates.
To better understand how D3 users currently implement interactions, we classified observed interactions in our corpus using a subset of Brehmer and Munzer's multi-level typology of visualization tasks~\cite{brehmer2013typology}. We observed 6 common interaction widget types: \textit{Brush, Click, Drag, Hover, Visualize}, and \textit{Zoom}. 52\% of the examples we classified contained at least one or more combinations of interaction widgets (e.g \textit{Hover, Click, Zoom}).

First, our findings show that interactions involving the selection of data points such as \textit{Hover} are used in the vast majority of interactive examples (62\%, see~\autoref{fig:int_types}). Second, some interactions seem to be coupled with specific visualization types; for example, D3 network graphs often support the drag interactions. Finally, the total interactions implemented varied by visualization type. For instance, network graphs tended to have at least one interaction whereas Line charts tended to have no interactions implemented. This suggests that in D3, \textit{interaction programming is dependent on the visualization design process, requiring support features that are both design-aware and code-aware.}

\subsection{Design Considerations}
\label{sec:analysis:design-considerations}
Based on these results and our personal experiences using D3, we provide three design considerations for Mirny.

\noindent\textbf{C1: Prioritize familiar interfaces/workflows.}
Past research highlights a need for tools to be integrated into users' workflows~\cite{drosos2020wrex,battle2022exploring} and familiar interfaces~\cite{gale2020sigma}. Furthermore, compatibility with other development platforms was one of the original design goals of D3~\cite{bostock2011d3}. For these reasons, we prioritize compatibility with users' existing workflows and preferred interfaces, reducing the burden of introducing yet another new tool to do their design prototyping process. Specifically, Mirny is designed to use either visualization templates \emph{or} existing code as input and makes it easy to export code so it can be used in other tools such as Observable~\cite{observable:2016}. We summarize the Mirny interface design and explain how it supports visualization prototyping in \autoref{sec:system-overview}.

\noindent\textbf{C2: Design adaptive recommendations to match current practice.}
Our prior work reveals that with the widespread implementation of D3 visualizations comes myriad conventions for when and how to implement interactions for specific D3 visualization types~\cite{bako2022streamlining}. To design effective interactions, D3 users have to navigate a complex space of existing design examples to identify compatible implementations.
By providing automated recommendations for compatible interactions (inspired by existing recommenders, e.g., \cite{gotz2009bdvr,vartak2017towards,shani2002mdp}), 
we can help users quickly prototype their envisioned designs, rather than forcing them to scour the internet for relevant examples. We present our recommendation engine to support interaction prototyping in \autoref{sec:recommendation-model}.

\noindent\textbf{C3: Support automated code augmentation.} Prior work~\cite{hogue2020searchingd3, battle2022exploring} has established code reuse as a prominent implementation strategy employed by D3 users. 
In our past work, we developed code templates for common data visualizations 
and interactions that D3 users implement~\cite{bako2022streamlining}. However, the templates alone are not enough as users may have to modify or add to the templates to realize their target designs. To facilitate template modification and customization, users need automated integration of template[s] into existing user code, which we refer to as \emph{code augmentation}. This way, a user can prototype multiple interactive visualization designs with a simple mouse click. We explain how we support automated code augmentation in \autoref{sec:augmenting_code}.

\section{Mirny System Overview}
\label{sec:system-overview}
\begin{figure}[tbp]
    \includegraphics[width=0.85\columnwidth]{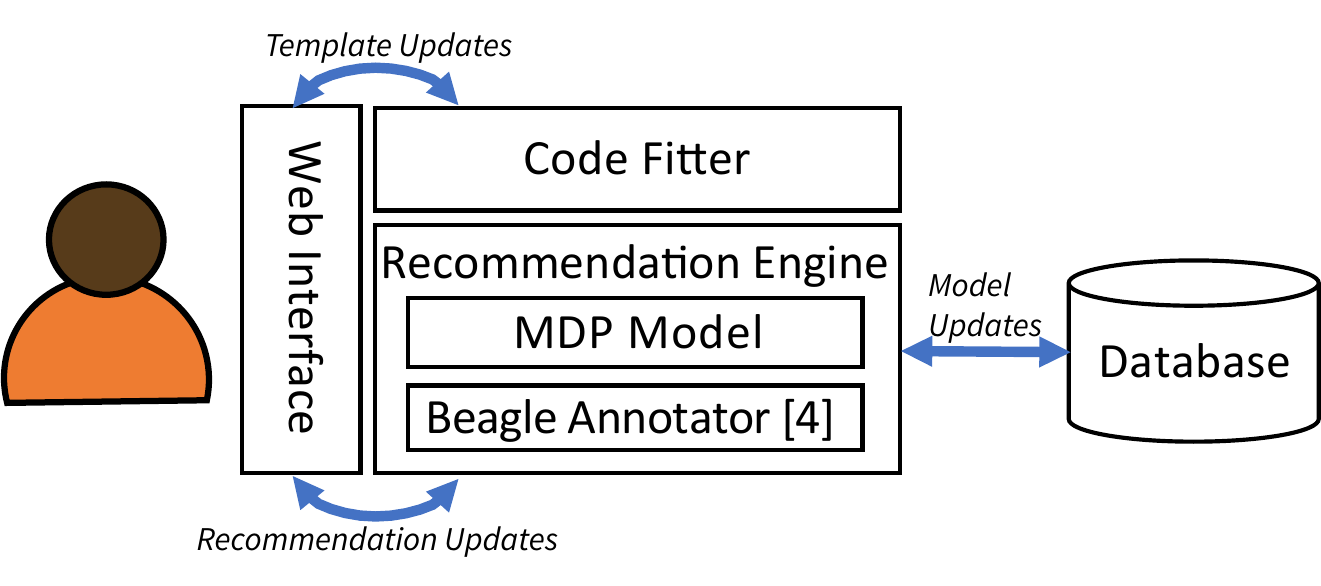}
    \vspace{-2mm}
    \caption{Mirny's architecture. The Mirny interface 
    captures user requests for interactions and recommendations. Using the Code Fitter, it generates the appropriate D3 specification and renders the results for the user via the Interface. Requests are also sent to the prediction model to receive recommended interactions for the user.}
    \Description{Mirny's architecture consists of a client-facing web interface through which users interact with the system. The web interface sends and receives template updates from the code fitter housed on the backend of the system. Recommendation updates are sent and received by the MDP model and the Beagle annotator which are both contained in the recommendation engine component on the backend. Finally, model updates are communicated between the recommendation engine and the database.}
    \label{fig:interfacearc}
    \vspace{-3mm}
\end{figure}

We implement two automated features in Mirny (recommendations and code augmentation) to facilitate interactive visualization prototyping in D3. These features are situated in a design-focused interface structured for integration with existing user workflows (addressing design considerations \textbf{C1} and \textbf{C2}). In this section, we describe Mirny's overall design (see \autoref{fig:interfacearc}).

The Mirny web interface provides a user with an interactive graphical user interface through which they can interact with the Mirny system.
The web interface consists of four core views:
\textbf{Templates Panel} (\autoref{fig:teaser}A), \textbf{Editor} (\autoref{fig:teaser}B), \textbf{Visualization Panel} (\autoref{fig:teaser}C) and the \textbf{Recommendation Panel} (\autoref{fig:teaser}D).

\textit{Templates Panel.} The Templates Panel is designed to help D3 users quickly generate an initial visualization prototype for a given dataset. Revisiting our example in \autoref{sec:motivating_scenario}: after Sandra uploads her dataset, she selects a Scatterplot from Mirny's list of templates. Mirny uses this template to identify attributes with compatible data types and generate an initial prototype visualization, which Sandra can view in the Editor (code) and Visualization Panel (output). Mirny supports templates for the 8 most popular visualization types identified in our prior work~\cite{bako2022streamlining}: Bar charts, Line charts, Scatterplots, Geographic Maps, Force Directed Graphs, Pie Charts, Donut charts, and Bubble plots. Mirny also leverages templates for 6 interaction types (Hover, Click, Brush, Drag, Zoom, and Visualize), which are displayed in the Recommendation Panel.

\textit{Editor.} The Mirny Editor is a live JavaScript editor tailored for prototyping interactive D3 visualizations. Given our focus on integration with existing user workflows, the Editor is designed in a similar style to complementary editors, such as those in Vega-Lite~\cite{satyanarayan2017vega-lite} and Observable~\cite{observable:2016}. For example, when Sandra modifies the code for her Scatterplot prototype in Mirny, it is as if she is making the change with similar editors, smoothing Mirny's learning curve.

\textit{Visualization Panel.} Mirny displays the visual output for the current prototype in the Visualization Panel. This panel updates live and is fully interactive, so when a user modifies the code, they can quickly test the results in the Visualization Panel. For example, when Sandra adds a Hover interaction to her Scatterplot, she can immediately test the output using this panel.

\textit{Recommendation Panel.} In parallel to the user's coding process, Mirny also recommends compatible interaction designs that could be added to the current prototype. These recommendations are displayed in the Recommendation Panel, ranked by their relevance to the target visualization type. Mirny also shows which interactions have already been implemented, such as Sandra's selected Hover interaction. The recommendations are powered by Mirny's recommendation engine, described in \autoref{sec:recommendation-model}. Users can click on any recommendation and it will automatically be integrated into their current prototype using Mirny's code augmentation feature, described in \autoref{sec:augmenting_code}.

\section{Recommending Code Snippets to Make D3 Visualizations Interactive}
\label{sec:recommendation-model}

\begin{figure*}[tb]
    \centering
    \begin{subfigure}[b]{0.6\textwidth}
         \centering
         \includegraphics[width=0.8\columnwidth]{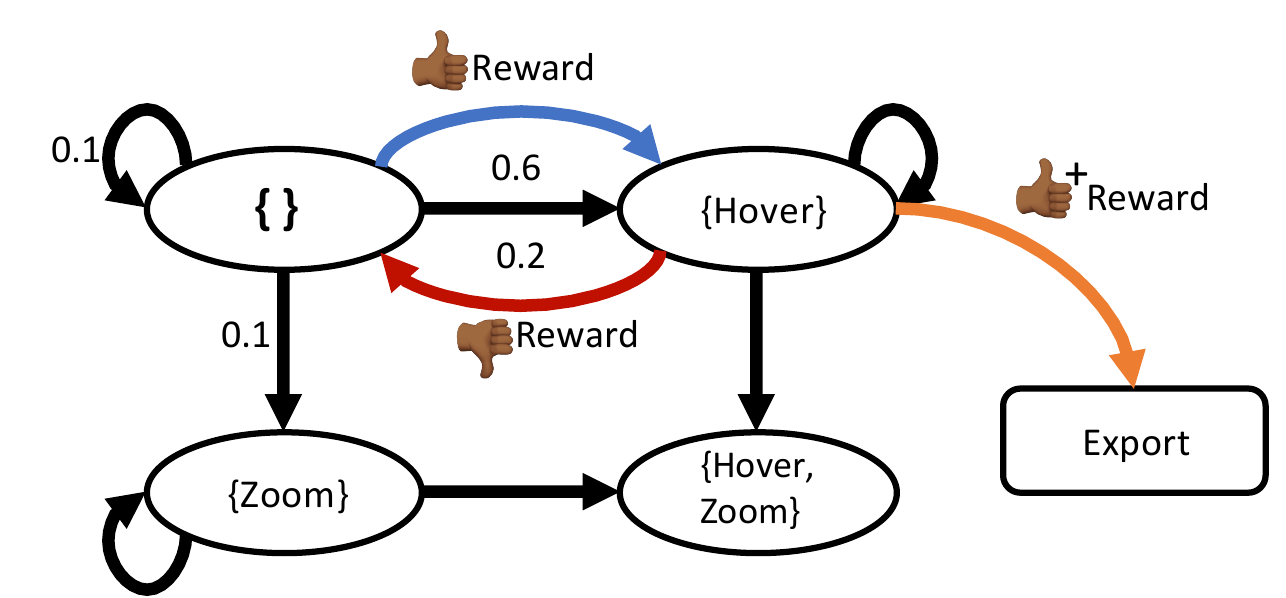}
        \caption{
        The Mirny recommendation engine models user interaction state using a Markov decision process (or MDP) model.
        Here is a simplified example of how the Mirny MDP model is structured with two interactions (zoom and hover), explained in \autoref{sec:recommendation:mdp-example}.
        }
        \label{fig:mdpexample}
        \Description{This figure depicts the MDP model that drives the Mirny recommendation engine. In the initial state, a user is presented with two possible interactions zoom and hover. From this initial state, there is an arrow to the left representing a 0.6\% probability that the user will transition to a new state where the hover interaction is implemented and will receive some positive reward. There is another emerging from the bottom representing a 0.1\% probability that the user will transition to a new state where the zoom interaction is implemented. There is also a 0.2\% probability that the user will transition to a hover state but undo their action to transition back to the initial state receiving a negative reward. There is also a 0.1\% probability that a user will remain in the initial state with no interactions. From the hover state a user can export their code or visualization where they will receive a large positive reward}
     \end{subfigure}
     \hfill
     \begin{subfigure}[b]{0.35\textwidth}
        \centering
        \includegraphics[width=\columnwidth]{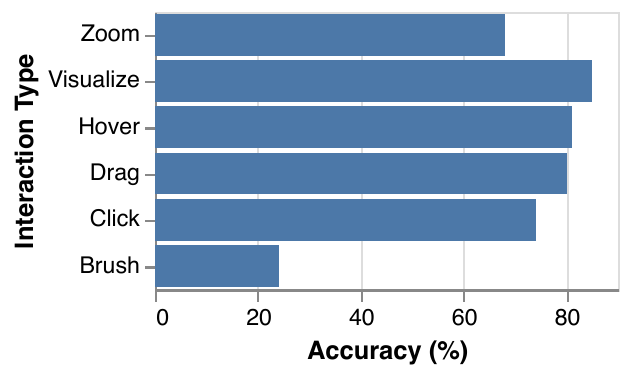}
        \caption{
        Results of the LOO cross-validation of the recommendation model across  per interaction type.
        }
        \label{fig:mdpaccuracy}
        \Description{This figure presents a bar chart of the results of the model evaluation across the 6 interaction types observed in our dataset.}
     \end{subfigure}
     \caption{Figure highlighting (a) an illustration of the behavior of the MDP recommendation engine and (b) the results of the evaluation of the recommendation model.}
\end{figure*}

Given the prevalent combinations of visualization and interaction types observed in our preliminary analysis (see \autoref{sec:prelim}), D3 users are clearly interested in \emph{interactive} visualizations. However, it may not be obvious to D3 users which interactions complement a given visualization. Revisiting our motivating example, Sandra may not know which interactions to implement to enhance her Scatterplot, such as Hover, Zoom, or Brush.

One approach to solving this problem is to \emph{automatically recommend compatible interactions} anchored on the implementation practices of past D3 users, for example, by training machine learning models to predict complementary interactions for a given visualization  (addressing design consideration \textbf{C2}).
However, users' programming behaviors and design preferences evolve as their interests shift over time~\cite{gotz2009bdvr,vartak2017towards}. Thus, a one-shot approach to model training risks generating stale results as users advance their visualization projects.
To keep up with the evolving nature of user preferences, our recommendations must also \textit{adapt} to shifts in user behavior over time. Hence, we leverage Markov Decision Process (MDP)~\cite{bel1957markov} to model users' interaction preferences and recommend complementary interactions to users as they program D3 visualizations. Here, we provide an overview of the prediction models that drive Mirny's recommendations.

\subsection{Predicting Which Interactions to Implement}
\noindent\textbf{Modeling Interaction States.}
To make recommendations, we first need an understanding of the current state of a user's visualization code and use that information to predict what interactions to recommend to the user. We define the MDP for our recommendation engine as a set of \textit{states S}, where each state \(s \in S\) is a set of the interactions that have already been implemented in the user's code \(\{i_1,i_2...i_n\}\).
Given that the user could pick any number of interactions to implement, we create a separate state for every possible combination of interactions that we observed from Bl.ocks.org.
Furthermore, this interaction set often starts out empty as D3 users seem to implement functional visualizations before incorporating interactions. For this reason, we make the empty state the initial state for our model.

For any given state $s \in S$,  our MDP model produces a ranked list of valid interaction recommendations, ordered by how likely they are to be implemented based on observed frequencies of interaction types among our D3 examples (see \autoref{sec:prelim}).

\noindent\textbf{Adjusting the Model Based on User Feedback.}
Given our interaction states, we model potential interaction implementation steps through a set of \textit{user actions A}, where each action $a \in A$ represents a user's reaction to a recommended interaction $i_r$. The user can react in one of four ways,
given the user's current state $s$, and the corresponding future state $s'$: 

\begin{itemize}[nosep]
    \item \textbf{Accept} the recommendation by clicking on the corresponding button in the Mirny interface (see \autoref{fig:teaser}D), thereby moving to state $s'$, given by the equation:
        \vspace{-1mm}
        \begin{equation}
            P[s, i_r, s'] = \frac{observations(s')}{observations(s)},
        \end{equation}
    where $observations$ computes the total observations of $s$ (i.e., the combination of interactions) in our corpus.
    \item \textbf{Export} the visualization code files (see \autoref{fig:teaser}E) and exiting Mirny. This is a cue that the user may be satisfied with her visualization and wants to save it for future use, calculated as:
        \vspace{-1mm}
        \begin{equation}
          P[s, i_r, export] = \frac{observations(s')}{\sum\limits_{s \in S}observations(export|s)}
        \end{equation}
    \item \textbf{Undo} an added interaction,
    suggesting the user was dissatisfied with the interaction and wants to remove it:
        \vspace{-1mm}
        \begin{equation}
          P[s, i_r, undo] = \frac{observations(s')}{\sum\limits_{s \in S}observations(undo|s)}
        \end{equation}
    \item \textbf{Ignore} all of Mirny's recommendations $I_r$ by not clicking on any of them in the interface, remaining in the same state $s$:
        \vspace{-1mm}
        \begin{equation}
            P[s, i_r, s] = 1 - P[s, i_r, export] - P[s, I_r, s'] -  P[s, i_r, undo]
        \end{equation}
\end{itemize}

Each time the user transitions to a different state, the probability distributions for outgoing transitions from $s$ will be recalculated to account for this additional information.
We normalize all corresponding probabilities to ensure that they sum to one.
Note that since the total interactions that a user will implement are relatively small, these distributions can be computed easily.\\  
\noindent\textbf{Predicting Transitions Between Interaction States.}
We model how likely the user is to transition from one state $s$ to another $s'$ as a transition probability between states. This is represented as a \textit{transition function T}, which computes the probability that an interaction $i$ recommended in state \(s_i \in S\) leads to state \(s_j \in S\), or $T(s_i |i_r, s_j)$. Given the user's current state $s \in S$, a recommended interaction $i_r$, and a corresponding future state $s' \in S$, we define our transition function probability as follows:
\vspace{-2mm}
\begin{equation}
    T_r(s, i_r, s') = P[s,i_r, s']
\vspace{-2mm}
\end{equation}
Note that an interaction will only be recommended if it has been observed before. However, we derive our states and transition probabilities using our observations from \autoref{sec:prelim}, ensuring that our MDP model covers D3 interactions that users often implement.

When a user does or does not choose one of Mirny's recommendations, the MDP is rewarded or penalized accordingly with a positive or negative \textit{reward} \(R\). We assume that the reward \(R\) for a user's reaction to recommendation $i_r$ to be: a small positive number if the user \textbf{accepts} $i_r$, a large positive value if the user \textbf{exports}, a large negative value if the user \textbf{undos}, and 0 if the user \textbf{ignores} $i_r$. In this way, the MDP can adapt to match users' behavior.

\subsection{MDP Example.}
\label{sec:recommendation:mdp-example}
We demonstrate the behavior of the recommendation engine using a simplified example based on our motivating scenario with Sandra from \autoref{sec:motivating_scenario}.
A simplified MDP model is depicted in \autoref{fig:mdpexample}.
In this example, we assume that there are only two possible interactions that Sandra will want to implement: zoom and hover.
With two interactions, there will be four possible states for our model, shown in \autoref{fig:mdpexample}: no interactions implemented (the initial state), zoom has been implemented, hover has been implemented, and all interactions have been implemented (the terminal state).

Assuming Sandra starts in the initial empty state with her Scatterplot, we can calculate initial transition probabilities, where four (fictitious) transition probabilities are provided in \autoref{fig:mdpexample}. The intuition behind these probabilities is as follows: Sandra may iterate on her Scatterplot before incorporating any interactions (represented with probability 0.1), and Sandra is much more likely to implement hover than zoom interactions (probabilities 0.6 and 0.1, respectively). There is also a possibility that Sandra could implement a hover interaction but then change her mind and click on the undo button (represented with probability 0.2). For the empty state, our model provides a ranked list with two items, where the first item is hover, and the second item is zoom.

In our motivating example, Sandra clicks on the first recommendation of hover, representing the \textbf{accept} reaction. In this case, the MDP model receives a small positive reward for providing a useful recommendation, represented by the blue line in \autoref{fig:mdpexample}. However, suppose that Sandra \textbf{undos} her last added interaction, this results in a negative reward to the  MDP model, represented by the red line in \autoref{fig:mdpexample}. If Sandra \textbf{exports} her visualization after adding the interaction, the MDP model receives a large positive reward for providing helpful recommendation[s] to Sandra.

\subsection{Preliminary Model Accuracy Results}
To ensure that our recommendations align with our observations from \autoref{sec:prelim}, we performed leave-one-out cross-validation~\cite {friedman2001elements} of the MDP model using our coded dataset. For each visualization example from our dataset, assuming the initial empty state, we classify a recommendation as correct only if they include all interactions implemented within that example. Overall, our model has an average accuracy of 76\% (see~\autoref{fig:mdpaccuracy}). For certain visualization types, this accuracy can be improved, e.g., for zooming interactions (68\% accuracy). We find that our model performs poorly for brush interactions due to the relatively small number of data points for this interaction type in our training data (see~\autoref{fig:data_results}). Such cases can be improved by adding more input observations for the model to train on. However, this evaluation does not take into account how users' reactions to recommendations may improve the model's predictions over time. We discuss our evaluation with actual D3 users in \autoref{sec:evaluation}.

\subsection{Predicting Visualization types}
We described how the  recommendation engine supports interaction predictions when the user selects a visualization template first. However, our implementation in the Mirny system can also support recommendations even when the user does not select a template. To do this, Mirny utilizes the Beagle Annotator developed by Battle et al. \cite{battle2018beagle} to enable real-time prediction of the visualization a user is currently implementing. With this annotator, Mirny can predict if a user is implementing a Scatterplot, or some other visualization type by extracting the SVG elements of the visualization and running it through the Beagle Annotator's decision tree classifiers.

\section{Augmenting User Code}
\label{sec:augmenting_code}
Recommending relevant interactions could help D3 users find and prototype interesting interactions for their visualization designs. However, these recommendations alone would not be very useful to visualization designers. Users would still have to search for example implementations and figure out how to integrate the code into their prototypes. A major goal of this work is \emph{to reduce the challenge of integrating code from different sources when prototyping visualization designs.}
We envision a user clicking just a single button to automatically add an interaction to her design specification (addressing design consideration \textbf{C3}).

The D3 API supports six interactions (i.e. \emph{Hover, Click, Brush, Drag, Visualize and Zoom}) which have been identified and analyzed in~\autoref{sec:prelim}. The underlying code base for these interactions has been manually extracted into pre-generated templates that Mirny uses to automatically generate code for selected interactions. Once a user selects a recommended interaction,
we apply a series of code rewriting steps to merge the template into the user's current code file. Specifically, we translate the user's code file into its corresponding abstract syntax trees (or ASTs), analyze the trees for potential insertion points, and merge the interaction code template with the user's code file at a selected insertion point. This process happens in four steps which we describe below.

\noindent\textbf{Step 1: Identify Input Variables for the Interaction Template.}
\label{sec:augmenting_code:identify_variables}
Interactions often result in changes to the state of the visualization. This may include displaying additional information about the selected data point through a tooltip or making modifications to the state of the selected mark (or non-selected marks) such as changing its position on a drag event. To create meaningful code from interaction templates, specific inputs from the user's code need to be extracted. However, given the abstract nature of code, it is initially unclear which variables in the user's code file map to the required inputs within a given interaction code template. Considering our running example from \autoref{sec:motivating_scenario}, if the hover interaction needs to update certain properties of selected circle marks in Sandra's Scatterplot, then the code that sets properties for the circle marks, and the data attributes passed as inputs to the mark properties, need to be incorporated into the hover template. 

Hence, the first step of the code augmentation process is to identify analogous variables between the user's code and the interaction template. Since the properties needed for the interaction templates are known beforehand, we can identify the appropriate data attributes in the user's code by parsing through its AST.

\begin{figure*}[tbp]
    \centering
   \includegraphics[width=0.9\linewidth, clip]{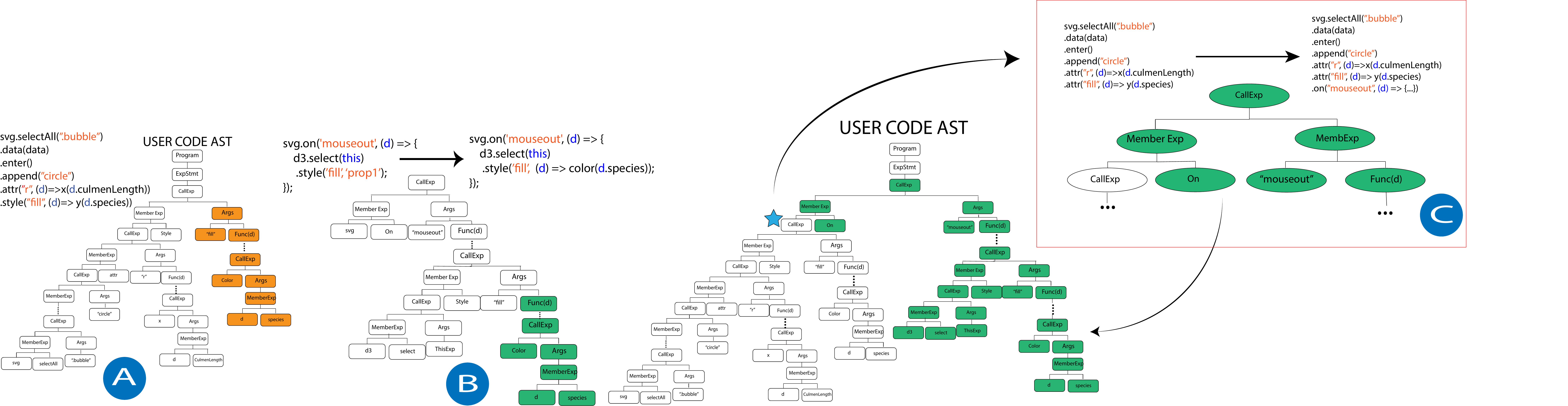}
  \caption{Given a recommended code snippet, e.g., a template for ``Hover'' interactions in D3, the Mirny system supports \emph{automatic code augmentation}, where the user's D3 code is mapped to an Abstract Syntax Tree (AST) that Mirny uses to populate and insert the code snippet. In (A), Mirny detects the ``fill'' variable in the user's current code (highlighted in orange). In (B), Mirny uses the ``fill'' variable to automatically populate the ``Hover'' template (highlighted in green). Finally, in (C), Mirny inserts the populated ``Hover'' template into the user's code at a detected anchor point in the AST (marked with a star).}
  \Description{This figure contains 3 code AST trees labeled A through C. Each AST represents a state in the code augmentation process.}
  \label{fig:code-augmentation}
  \vspace{-3mm}
\end{figure*}

Using our motivating example, when Sandra selects the \textit{Hover} interaction from the recommendation panel, our templates assume that at the end of the \textit{Hover} event, properties like the color and radius of the circle mark would need to be restored to its original value.
The Code Fitter parses through Sandra's code to identify the input variables that are needed to make a viable template, searching within API calls, variables, and function declarations for the needed components. For instance, suppose Sandra's code creates and sets the fill of the circle marks in a scatter plot using the following code:

\begin{lstlisting}
svg.selectAll('.bubble')
	.data(data)
	.enter().append('circle')
	...
	.attr('r', (d) => radius(d.culmenLength);)
	.style('fill', (d) => color(d.species););
\end{lstlisting}

The code fitter parses through the AST of her code to find where the circle marks' \texttt{fill} attribute is assigned and extract the assigned value (or in this case function) (see~\autoref{fig:code-augmentation}a). 

\noindent\textbf{Step 2: Populate the Interaction Template.}
\label{sec:augmenting_code:fitting_template}
Once all the required input variables have been identified, the interaction template is modified to include these input variables. The generic interaction template contains \texttt{prop} variables that are replaced with the appropriate input variables from the user's code. For instance, our default \texttt{Hover} template changes all marks to a template-specified color. At the end of the hover interaction, this color will need to be reset. Thus all API calls are updated  with the right variable names and parameters as needed. In our running example, the \texttt{mouseout} event of the hover interaction will be populated with the original function to determine the circle's \texttt{fill} as seen in~\autoref{fig:code-augmentation}b. If an attribute is not found within the code (see description in step 1), a default value is assigned. However, designers have the discretion to change the default values our templates provide after the code augmentation process is completed.\\

\begin{imageonly}
\begin{minipage}[]{0.4\columnwidth}
\begin{lstlisting}[title = Generic Template]
.on('mouseout', (d) => {
    d3.select(this)
     .style("fill", "prop1");
});
\end{lstlisting}
\end{minipage}%
\hfill
\vrule
\hfill
\begin{minipage}[]{0.5\columnwidth}
\vspace{0.5mm}
\begin{lstlisting}[title = Populated Template]
 .on('mouseout', (d) => {
    d3.select(this)
     .style("fill", (d) => color(d.species));
});
\end{lstlisting}
\end{minipage}
\end{imageonly}

\vspace{1mm}
\noindent\textbf{Step 3: Identify Anchor Points in the Users' code for Insertion.}
\label{sec:augmenting_code:anchor_points}
In D3, the specification of interactions is \emph{imperative}~\cite{satyanarayan2017vega-lite}, hence, the interaction code cannot be placed just anywhere within an existing D3 specification. Before modifying the user's code, we must first identify appropriate locations within the code to include new interactions. We refer to these locations as \emph{anchor points}.

Anchor points signify nodes within the AST that need to be changed to support a new interaction. For example, to add the Hover interaction to Sandra's code, a new node that maps to the interaction code has to be created and added to her code at the point where the visual marks are created. Because we know she is creating a Scatterplot, this would translate to the code block that contains the \texttt{.append(circle)} API call. ASTs are traversed using a depth-first approach, hence, identifying the anchor point simply involves iterating over the AST, with the objective of finding the node which has the appropriate data binding API call in its child branch. For our hover example, the anchor point would translate to the first node we identify that has the \texttt{.append(circle)} expression as a member of its child nodes as seen in~\autoref{fig:code-augmentation}c. 

Anchor points for most visualizations generally can be found where the data binding to visual marks occurs. However, for complex interactions such as \emph{Zoom} or \emph{Brush} interactions helper functions and new visual elements need to be added to the user's code, as such multiple anchor points are needed. Helper functions need to be accessed from anywhere in the D3 script (i.e. have a global scope), hence their anchor point is by default the body of the entire script. Elements, such as \texttt{rects}, to enable zooming or brushing over sections of the visualization need to be added over the entire SVG, as such they use the node that binds the \texttt{<SVG>} element to the web page, as their anchor point.

\noindent\textbf{Step 4: Insert the Populated Template into the User's Code.}
\label{sec:augmenting_code:updating_code}
Once a template has been populated and the anchor point has been found, the template code is then added as a new node in the user code at the selected anchor point. As a result, the populated template code becomes a new sub-tree within the AST of the user's code. The template's root node can either be \emph{appended} or \emph{pre-pended} to and in some cases \emph{replace} the anchor point.

The nature of the template node and the semantic correctness of the code are taken into consideration when determining which of the insertion actions to perform. Code that introduces a new code block after the anchor point typically will use an \emph{append} operation, whereas new code that goes before the anchor point will use a \emph{pre-pend} operation. For instance, adding new code that defines a tooltip for the Hover interaction will be included before we call any hover event. As such, the code for the tooltip will be added before the anchor point using a \emph{pre-pend} operation.
The \emph{replace} operation is usually used when adding new code as part of an already existing code sequence in the user's code. This is due to the fact that ASTs are traditionally traversed depth-first, so the new code will have to become a parent node to the anchor point. In our motivating example for the \textit{Hover} interaction, the code that handles the \texttt{mouseover} and \texttt{mouseout} events will need to be added to the entire anchor point's branch. This will require a \emph{replace} operation to make the template code's AST a parent node of the anchor point as seen in \autoref{fig:code-augmentation}c. Finally, Mirny converts the new AST back into code and updates the Editor to display the updated code file. 
\begin{lstlisting}
svg.selectAll('.bubble')
	.data(data)
	.enter().append('circle')
	...
	.attr('r', (d) => radius(d.culmenLength);)
	.style('fill', (d) => color(d.species););
	.on('mouseover', (d) => { ...});
	.on('mouseout', (d) => {...});
\end{lstlisting}

\vspace{1mm}
\noindent\textbf{Limitations with Babel.} The code augmentation feature helps users add interactions to their live D3 code in a single click. However, the code fitter uses the Babel compiler~\cite{babel2020} to transform D3 code into its corresponding AST, as a result, there is a limitation to its operation. The Babel compiler requires that the code to be transformed is both semantically and syntactically correct. Hence, the code augmentation feature will not work if a user selects an interaction on incorrect D3 code. In these cases, Mirny will alert the user about the issues and not complete the code augmentation process.

\section{Evaluation: User Study Design} \label{sec:evaluation}

In previous sections, we introduced two automated features with the aim of supporting template-based visualization prototyping. We implemented these features in a design-focused environment called Mirny. In this section, we focus on evaluating if these automated features lead to measurable improvements in how D3 users prototype interactive visualizations~\cite{ren2018reflecting,satyanarayan2019critical}. Our study was preregistered on aspredicted.org\footnote{Pre-Registration available at \textcolor{blue}{\protect\url{https://aspredicted.org/t2hb3.pdf}}}.
 
\noindent\textbf{Participants.}
    We recruited 20 participants from mailing lists and social networking sites (4 female, 15 male, and 1 non-binary) between the ages of 18 to 44. To ensure that all participants had a sufficient understanding of D3 to complete the study tasks under time constraints, we required participants to have at least 3 months of experience working with D3.js. 11 participants had 3 months to 1 year of experience, 3 had 2 - 3 years of experience, 4 had 4 - 5 years of experience and 2 had 5+ years of experience. Participants self-report being proficient in creating visualizations. Each session took approximately 1 hour and 30 minutes and participants were compensated with a \$20 Amazon gift card. 
    
\noindent\textbf{Tasks.}
    To observe how participants use these automated features in various visualization design contexts, participants were asked to perform two tasks for the study~\cite{wongsuphasawat2016voyager, ren2018reflecting}. Participants were allowed to create whatever visualization or interaction type they chose as long as they met the requirements for each task.
    \begin{itemize}[nosep]
        \item \textit{Task One: Targeted visualization design:} For this task, participants were asked to use the Cherry Blossoms dataset to create a visualization meeting specific design requirements. The output visualization[s] had to meet two requirements: target users of the output visualization should be able to  (a) identify correlations between two specific variables, and (b) interact with the data points in the visualization.
        
        \item\textit{Task Two: Exploratory visualization design:} Participants were asked to choose between the popular Cars\footnote{\textcolor{blue}{ \url{https://www.kaggle.com/abineshkumark/carsdata}}}and Iris~\cite{andersen1935irises} datasets, and then brainstorm what type of visualization they would like to create. Then, participants were asked to program this visualization and include at least one interaction in their visualization.
    \end{itemize}
   
\noindent\textbf{Experiment Conditions.} Participants experienced two different Mirny setups across our tasks, resulting in four total experiment conditions ($2\:tasks \times 2\:interface\:setups$). In one setup (baseline), participants only had access to Mirny's visualization templates, but not Mirny's interaction recommendations. In the second setup, participants had access to all of Mirny's features. Condition order was counterbalanced across the visualization tasks. We tuned our experiment conditions through a pilot study, described below.

\noindent\textbf{Pilot.}
    We conducted a pilot study to fine-tune our initial protocol. One important parameter we sought to tune was task difficulty, i.e., should participants be asked to create D3 visualizations from scratch as a task condition (the common case), or should participants be provided with Mirny's templates? In the pilot study, all participants (2 female, 1 male) were familiarized with Mirny and were asked to perform the tasks described above. For task one, we provided access to Mirny's visualization templates for two of the three pilot participants, while one participant created visualizations from scratch. Participants were able to use both the predefined templates and recommendations for task 2.
    
    We observed that the participants who were provided with templates only used Mirny's templates while the third participant copied code from a D3 example they found online. The participant who had to write their D3 code from scratch spent 45 minutes programming a single visualization and was unable to implement any interactions, thus unable to complete the task. The participants who were provided with templates took less than 15 minutes to create a single interactive visualization for the same task. These results show that it would have been difficult to conduct a one-hour experiment if participants had to create visualizations from scratch, and the experiment time could be drastically reduced if templates were always provided. Hence, we decided to provide templates for all tasks and experiment conditions. However, users could always opt to write their own code from scratch if desired.
     
\noindent\textbf{Protocol.}
    Participants were required to complete a demographic survey to determine eligibility for the study. Each session started with a Pre-study session where participants were familiarized with the study and signed the consent form. They then proceeded to a training session where the experimenter walked them through the interface functionality via a prerecorded demo of the interface. Participants were then given 5 minutes to explore the full Mirny interface themselves and ask any questions they had about the interface. Once participants were comfortable with the interface, they proceeded to complete our two study tasks. The experimenter limited guidance to only situations where the user reached a complete standstill and was unable to figure out the next step. Participants were allowed to use any external resource they wanted such as Stack Overflow or Bl.ocks.org. No time limits were placed for any of the tasks. Once all the tasks were completed, the participants completed a short exit survey and interview. 
    
\noindent\textbf{Data Collection}
    For each session, an experimenter observed the participants while they performed their tasks and took notes. Logs of participants' interactions with Mirny were also recorded through the Mirny back-end, which captured all click and keyboard events. We collected data from the post-task survey on participants' ratings of Mirny's features to capture participant feedback. Finally, video and audio were recorded to capture participants' interactions with the interface and responses to the exit interview questions.

\begin{figure*}[tbp]
\begin{subfigure}[h]{0.25\linewidth}
    \centering
    \includegraphics[width=0.8\textwidth, clip]{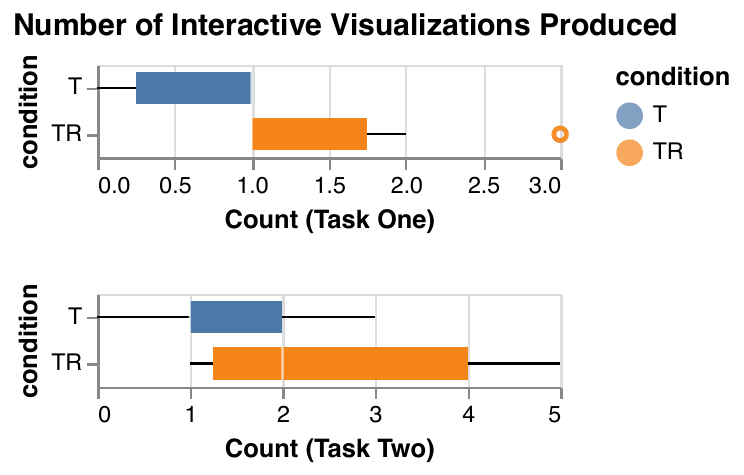}
    \caption{Count of interactive visualizations produced by participants.}
    \label{fig:interaction_count}
    \Description{Box plots highlighting the number of interactive visualizations produced by participants in the user study. Participants in the templates + recommendations condition for both task one and task two have higher number of interactive visualizations}
\end{subfigure}%
\hspace{1em}%
\begin{subfigure}[h]{0.25\linewidth}
    \centering
    \includegraphics[width=0.9\textwidth, clip]{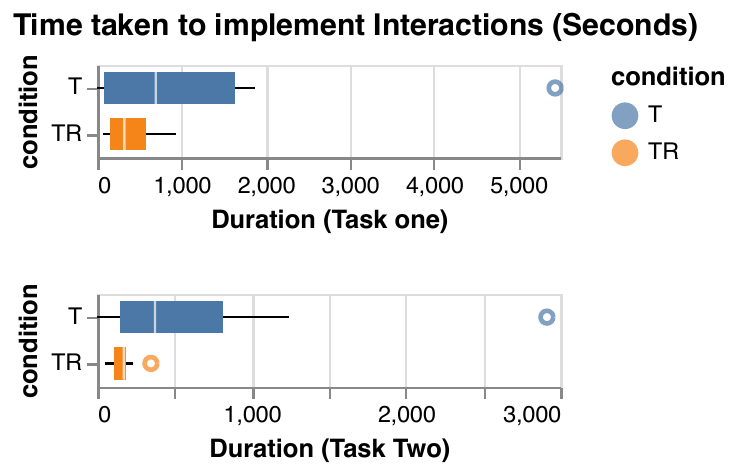}
    \caption{Distribution of time taken to program a single interaction.}
    \label{fig:interaction_time}
    \Description{Box plots highlighting the time taken to implement  interactions in a visualization by participants. Participants in the templates + recommendations condition for both task one and task two, spend less time creating interactions. There is an outlier in the templates-only condition that spent upwards of an hour creating interactions for task one.}
\end{subfigure}%
\hspace{1em}%
\begin{subfigure}[h]{0.4\linewidth}
    \centering
    \includegraphics[width=.9\textwidth, clip]{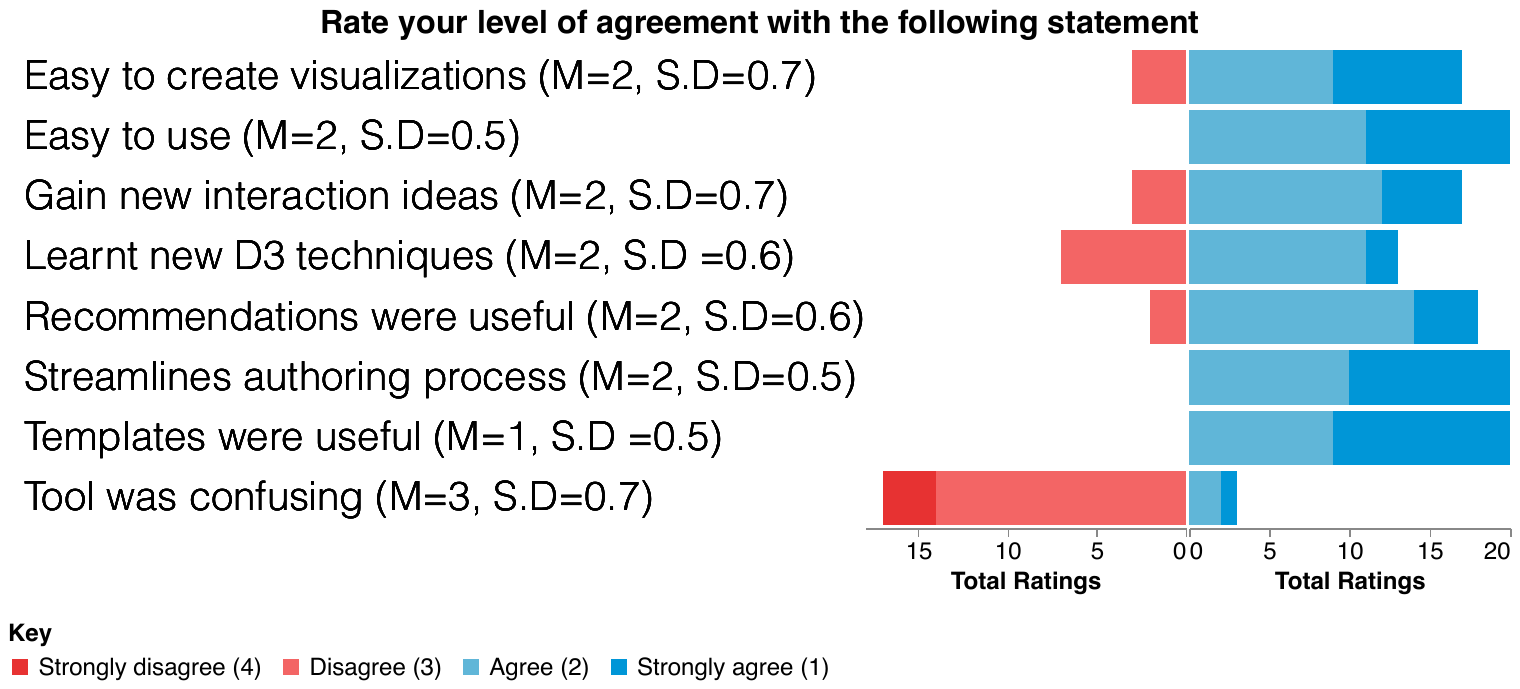}
    \vspace{-2mm}
     \caption{Distribution of post-study questionnaire responses.}
    \label{fig:surveyresponses}
    \Description{Diverging stacked bar chart showing participants' responses to exit survey. Participants found visualizations easy to create using Mirny and were able to learn new D3 techniques.}
\end{subfigure}
\caption{Results from analysis of data collected during evaluation of Mirny.  T = Templates only, TR = Templates + Recommendations.}
\vspace{-3mm}
\end{figure*}

\section{Evaluation: Analysis \& Results}
\label{sec:results}

 Here we present three analyses of our study data:
 measuring session duration and design prototypes[s] (\autoref{sec:results:automation}), classifying template usage and modification (\autoref{sec:results:templates}), and summarizing participant feedback (\autoref{sec:results:feedback}).
    
\subsection{Measuring Session Duration and Design Prototypes}
\label{sec:results:automation}

We formed two hypotheses to test how Mirny's code recommendation and augmentation support the visualization prototyping process:  
 \begin{itemize}[nosep]
        \item \textbf{H1}: Users produce a higher number of interactive visualizations provided with visualization templates and interaction recommendations, versus only visualization templates.
        \item\textbf{H2}: Users spend less time implementing interactive visualizations when provided with visualization templates and interaction recommendations, compared to when users only have access to visualization templates.
    \end{itemize} 
    
To test our hypotheses, we collected data on the number of visualizations created, the number of interactions participants implemented for each visualization (\textbf{H1}), and the time taken to implement a fully interactive visualization, i.e., time taken to render a single visualization + time taken to implement a single interaction (\textbf{H2}). We fit linear mixed effects models~\cite{barr2013random} with the condition and condition order as fixed effects and participant and dataset as random effects. 

Based on the results of our pilots, templates were provided for all participants. As a result, we did not statistically test the number of visualizations explored by each participant since all participants used the templates. On average participants created 2 visualizations for each design task. We find a significant effect for interface layout on the number of interactions [\textbf{H1}: \(({\chi}^2(3, N=20) = 13.406, p=0.011)\)] and time to implement an interaction [\textbf{H2}: \(({\chi}^2(1, N=20) = 3.85, p=0.0499)\)]. Recommendations + code augmentation led to \emph{1.9 more interactions implemented per visualization} compared to only 0.7 interactions when using only templates. Participants also \emph{implemented interactions 3.2x faster} when recommendations and code augmentation were provided (5 minutes versus 16 minutes for templates only). Our survey results show that participants found the recommendations to be useful(Median(M)=2/Agree, $\sigma=0.6$) and instrumental in helping them gain new ideas of interactions to implement (Median(M)=2/Agree, $\sigma=0.6$).

\subsection{Classifying Template Usage and Modification}
\label{sec:results:templates}

To assess the efficacy of our automated features in helping users program visualizations, we formed two hypotheses:

 \begin{itemize}[nosep]
        \item \textbf{H3}: Users produce a higher number of executable visualizations when provided with visualization templates compared to when users program visualizations from scratch.
        \item \textbf{H4}: Users find D3 more interesting and more useful when provided with both visualization templates and interaction recommendations, versus only visualization templates.
    \end{itemize} 
 
However, based on the results of our pilot study we needed to provide templates for all participants,
affecting our analysis methods. While we do not quantitatively test \textbf{H3}, we have informal support for \textbf{H4} based on our observations of user behavior and feedback from participants.

\noindent \textbf{Participants default to templates to complete tasks.} We expected that some participants would want to write their own code and provided an option to support this. Contrary to our expectations, \emph{every participant in the study} defaulted to using Mirny's templates, regardless of D3 expertise. Most participants settled on Scatterplots as their final design. We suspect that some participants may have been influenced by the word `correlation' in task one which primed participants to choose scatterplots as their final design. Unfortunately, this behavior was unintentional and the fixation on scatterplots could have been mitigated by a more careful examination of the task description. Nevertheless, participants often explored other chart types such as line charts, bar charts, and pie charts before deciding what visualization to use.

\noindent\textbf{Template Modification Strategies.} Our findings amplify prior observations that templates are a great starting point for users but require modifications to fit users' needs. We observed a total of \emph{33 distinct code modifications} to Mirny's visualization templates (n=22) and interaction templates (n=11). The majority of these modifications involved changes to accommodate \textit{stylistic preferences} of participants (n=22), such as changes to the fills for marks and their associated color schemes, changes to the displayed layout of visual elements such as text, or changes to add new behavior to interactions. For example, P16 modified the code for a hover interaction to remove the default tooltip and instead change the opacity and size of other circles outside of the \texttt{species} class of the selected data point(s). P16 also modified the code to increase the size of all circles in the selected species class to make them more prominent.  

\begin{figure}[tbp]
    \centering
    \includegraphics[width=0.9\columnwidth, clip, trim=0cm 0cm 0cm 0cm]{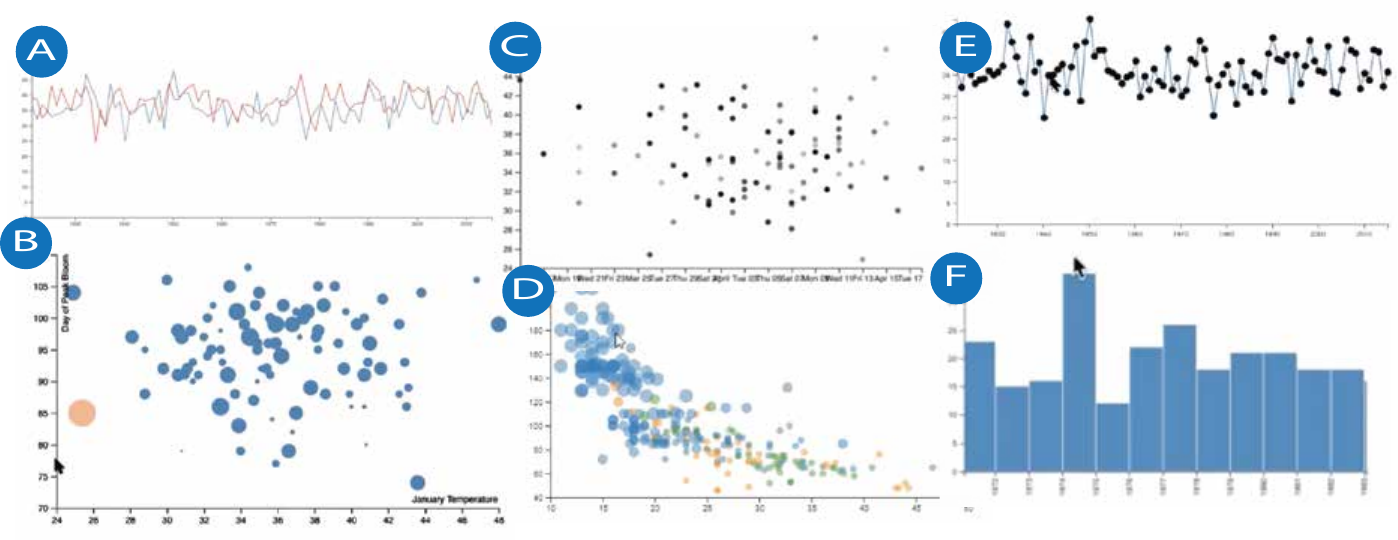}
    \vspace{-2mm}
     \caption{Examples of visualization designs created by participants during the experiment. (A), (B), (C), and (E) are visualizations of the Cherry Blossoms dataset  depicting different attributes in the dataset such as  yearly records of January Snow and February Snow (A),  day of peak bloom vs January Temperature (B), etc. (D) and (F) are visualizations of attributes from the cars dataset.}
     \Description{A collage of 6 visualization designs consisting of a line chart, a bar chart, and four scatterplots. Each visualization is labeled right through left from A through F.}
    \label{fig:designexample}
     \vspace{-3mm}
\end{figure}

We observed 6 occurrences of participants modifying Mirny's visualization templates to \textit{transform data} such as aggregating, sorting, or grouping data attributes. For instance, P15 modified the bar chart template to aggregate the number of cars produced each year and group them by country of origin. We also observed 5 modifications to \textit{introduce new data encodings} or visual marks to the templates, which resulted in particularly interesting visualization redesigns. For example, P19 introduced an encoding mapping the \texttt{LastFrost} attribute to circle size in their Scatterplot implementation; more examples can be seen in~\autoref{fig:designexample}. 

\noindent\textbf{Unexpected uses of templates.} 
We observed an unexpected use case for templates by some participants while completing the design tasks. A small subset of our participants (n=4) used an initial template to \textit{perform exploratory data analysis}, then chose a different template later on to present the results. For instance, when completing task two, P7 used the bar chart template to analyze the distributions of various data attributes. Then, they used the resulting insights to select attributes to present in their final Scatterplot. 

\subsection{Participant Feedback}
\label{sec:results:feedback}
Participants had varied reactions to the automated features provided in the Mirny system. In particular, participants often explained that the templates helped them integrate their design thinking into their programming process because Mirny provided starter code which helped save time finding suitable code. Participants mentioned that on their own, they would have to spend hours finding example code, which is known to be a major design challenge for D3 users~\cite{battle2022exploring} and even visualization designers in general~\cite{bako2022understanding}. For instance, P11 comments \textit{``those [templates] are pretty valuable because there's always that boilerplate of code you would have to write if you don't have those templates.''} Similarly, P10 said \textit{``I think it's good that it's smart enough to do this. And normally, doing these interactions is not easy. For example, I wouldn't know how to implement brushing. I know, it's a very common thing. But I believe every time I want to do it, I have to search for it online.''} Participants mentioned that having access to the code for both the visualizations and interactions helped them to  improve their understanding of the corresponding D3 designs.

The ability to quickly iterate over different interaction prototypes also allowed them to weigh their design options to understand and identify which designs were suitable for the task. This feature also helped participants to avoid the sunken cost fallacy that would have otherwise made them stick to a specific design only because of the time they had invested in creating their design. P20 states \textit{``...instead of having to look for separate things myself, all these recommendations are in one place so it's very easy to try out which interaction is suitable for which kind of visualization. So I just click and then I can undo and then be like, okay, I probably don't want this. Whereas if I were to implement them myself, I would want to use something that I made, because I already spent time on it. So I want to use it, even though I probably shouldn't.}

More experienced participants explained that access to code may not be useful for them due to their existing knowledge of D3, but also agreed that an environment similar to what Mirny provides would be useful for novices who are just starting to learn D3. For example, P9 instructs a visualization class and mentions that teaching D3 is a difficult task as students typically need to learn HTML, CSS, JavaScript, and D3 all at the same time, hindering students' ability to prototype different visualization designs to understand their perceptual pros and cons. \textit{``I also teach people d3. And I think this would be great...Having something like this, that creates the chart and the code and comments the code, so you can read the code. They [students] would like cry tears of joy to have this."}
    
Participants welcomed the automated features demonstrated in Mirny, however, our approach is not perfect, and participants made suggestions for further refinement. Participants expressed the need for insight on why certain recommendations are made, e.g., P10 said \textit{``instead of having the recommendation shift, just give more insight as to why this is being recommended, basically"}. Participants also suggested including more widgets to support further customization of D3 code, e.g., P15 said \textit{``...maybe adding something like the components, to support selecting an axis or legend. If those are also given as part of templates to completely customize code, that might be more helpful"}. Participants also found the live updates feature provided in Mirny to be a hindrance when debugging their code. Participants suggested adding a run button to allow them to update the visualization output when desired.

\section{Discussion and Future Work}
 In this work, we present an approach to visualization prototyping that relies on template-based recommendation and augmentation. Our results demonstrate that our approach allows users to generate multiple visualization and interaction designs with significantly less effort and time. Our evaluation highlights strategies that participants engage in to modify templates to fit their design goals. Participants' feedback also shows that our approach saves users valuable time that they would have otherwise spent seeking design implementations; allowing them to  invest their time exploring and evaluating design alternatives. We reflect on the implications of our findings and future directions for research on understanding visualization design, supporting user creativity and autonomy as well as advancing template-driven design.

\textbf{Understanding visualization design reasoning.} Visualization designers often start building their designs from an example or template~\cite{bako2022understanding}, however, these templates are just a starting point for further design exploration. We find that when provided with visualization templates, participants would often make changes to implement their desired design ideas. We find that the key reasons why participants \emph{augment our templates} are to: (1) accommodate their stylistic preferences, (2) perform data transformations, and (3) introduce new encoding[s] to the visualization. Modifying templates or examples is a common phenomenon amongst designers, yet the visualization community has hardly explored how designers build on and source ideas for their visual designs. In the future, we need to deeply investigate and understand the designer's reasoning during the design process to understand how the context of use and design tasks influence the changes a user makes to design a template. Furthermore, more research is needed to understand when and how code augmentation fits into the visualization prototyping process. Finally, research also needs to explore the role datasets and stylistic preferences play in the visualization design process. 

\textbf{Balancing automation and user agency in visualization generation.} The recent rise in program synthesis tools like GitHub's Copilot has made it possible to automatically generate code~\cite{chen2021evaluating}. However, as with all automated systems, users lose control over the code specifications that are generated as these tools often generate buggy or outdated code that a user neither understands nor knows how to fix~\cite{asare2022is,dakhel2022github}; a limitation acknowledged in Copilot's documentation~\cite{github2023CopilotDoc}. While our approach relies on the same principles of automated code generation, its code generation process is built on manually curated code templates that incorporate domain-relevant expertise, safeguarding users from some limitations of automatic code generators. However, we acknowledge that the manual nature of Mirny's template curation limits its scalability to more examples. Future work could explore ensemble strategies that incorporate the strengths of both approaches. For instance, one could integrate Mirny's domain-specific data on D3 visualizations and its recommendation engine with Copilot to learn from a wider array of visualization and general programming examples to generate recommendations.

Visualization design is still a very subjective process that requires human input and creativity, evidenced by our participants' desire to modify the generated visualization designs. Current visualization recommenders (e.g., Show Me~\cite{mackinlay2007showme}, Voyager~\cite{wongsuphasawat2016voyager, wongsuphasawat2017voyager}, etc.), often treat system-generated designs as final output with little leeway (if any) for modifying the recommended designs, limiting user creativity. In the future, research needs to explore how automated visualization authoring tools can simultaneously preserve user agency and individual design preferences in the designs they generate or recommend~\cite{baorecommendations2022}. A potential approach is to explore how users can be incorporated into the design generation process providing an opportunity for the user and recommenders to co-design data visualizations.

\textbf{Modular visualization construction.} Inspiration for visualization designs often comes from multiple sources and designers often piece together components from these sources into their designs~\cite{bako2022understanding, zhang2020dataquilt}. Participants like P15 in our study also expressed interest in templates that break visualization elements into modular components to support further customization. This request echoes the sentiment of other D3 users shared online~\cite{battle2022exploring}. Future research on the use of design templates should allow for dialog between users, templates, and authoring tools to improve the visualization programming experience. For instance, modularized templates for each part of a visualization (e.g axes, marks) could allow users to mix and match designs. 

\textbf{Supporting complex template transformations.} We acknowledge that our choice to focus on only the common D3 visualization types and code integration for interactions is modest. However, this work is only a starting point in supporting dynamic template transformations for visualization design. We observed during our user study that users may be interested in changing a template from one visualization form to another. For instance changing a line chart to a connected scatter plot (see~\autoref{fig:designexample}b). An important direction for future work is to explore extending our automated features to support more complex template transformations such as changing the structure and arrangement of visual marks in response to data transformations. For instance, how can we dynamically transform a template that implements a bar chart to a stacked bar chart when a user performs data aggregation?

\section{Conclusion}
\label{sec:conclusion}
In this paper, we present a new approach to supporting visualization prototyping when using complex toolkits like D3. We contribute two template-driven automated features to help users program interactive D3 visualizations with less time and effort: (1) a \emph{recommendation engine} for suggesting complementary interactions to add to a D3 visualization, and (2) \emph{automatic code augmentation}  to incorporate suggested code snippets to live user code, even when users are \emph{not} following a pre-defined template. We demonstrate these features in Mirny and in a user study with 20 D3 users, we find that they enable users to create interactive visualizations faster and in fewer iterations compared to the typical D3 development workflow (i.e., when automated features are not available). We also observe three key strategies used by D3 users to customize visualization templates and unexpected uses of visualization templates during design. Our work highlights important takeaways on the use of dynamic templates and D3 users' perspectives that we hope will influence future work on visualization prototyping.


\begin{acks} 
We would like to thank members of the UMD Human-Computer Interaction Lab, the Human Data Interaction Group, the UW Interactive Data Lab, and our reviewers for their invaluable feedback. This work was supported in part by NSF award \# IIS-1850115 and a VMWare Early Career Faculty Grant. 
\end{acks}


\balance
\bibliographystyle{ACM-Reference-Format}
\bibliography{references}

\end{document}